\documentclass[prd,aps,a4paper,nofootinbib]{revtex4}  

\newif\ifusesec
\usesectrue  
   
\usepackage{graphicx} 
\usepackage{mathrsfs}
\usepackage{amsmath,amsfonts,amssymb}
\usepackage{multirow}

\newcommand{\beq}{\begin{equation}}
\newcommand{\eeq}{\end{equation}}
\newcommand{\bea}{\begin{eqnarray}}
\newcommand{\eea}{\end{eqnarray}}

\begin{document}

\title{Binary black hole scattering in the extreme-mass-ratio limit: time-domain waveform}

\author{Andrea Geralico}
\affiliation{
Istituto per le Applicazioni del Calcolo ``M. Picone,'' CNR, I-00185 Rome, Italy
}

\date{\today}

\begin{abstract}
The time-domain gravitational waveform emitted during the scattering of two nonspinning compact bodies is computed in the extreme-mass-ratio limit.
It provides a time-domain description of the gravitational radiation emitted during the encounter and complements previous first-order self-force calculations, which have primarily been formulated in the frequency domain. In this way, the present results provide a complementary representation of the radiative dynamics and facilitate a direct comparison between time-domain and frequency-domain approaches to the gravitational-wave signal.
The waveform is accurate to the fifth post-Minkowskian (three-loop) level and seventh post-Newtonian order, and will serve as a benchmark for future calculations by other methods, to first order in the mass ratio.
The results have been already tested in a previous work by constructing the energy and angular momentum fluxes, thereby computing the corresponding radiative losses at the 5PM and 4PM level, respectively, with the same PN accuracy, which are in agreement with recent amplitude-based calculations. 
\end{abstract}


\maketitle

\section{Introduction}

The increasing sensitivity of gravitational-wave (GW) detectors, together with the growing number of observed compact-binary coalescence events, has placed increasingly stringent demands on the accuracy of theoretical waveform models \cite{KAGRA:2021vkt,LIGOScientific:2026ifv}. Accurate waveform predictions are essential for both the detection and parameter estimation of compact-binary signals, as systematic waveform errors can bias the extraction of the physical parameters of the source (see, e.g., Ref. \cite{Christensen:2022bxb}). In particular, the accumulated phase of the waveform plays a crucial role in matched-filtering analyses and must be modeled with high accuracy over the long inspiral phase \cite{Jan:2023raq,Dhani:2024jja}.

In this context, a time-domain description of the waveform provides a particularly useful connection between the observable GW signal and the underlying binary dynamics. The orbital motion, radiation-reaction effects, and multipolar content of the emitted gravitational radiation evolve dynamically throughout the encounter, with the resulting waveform determined by their coupled evolution. A time-domain formulation therefore provides a natural framework for incorporating conservative and dissipative corrections to the orbital dynamics and for propagating these corrections directly into the emitted waveform. 

This perspective is especially relevant in perturbative approaches to the two-body problem. 
Post-Newtonian (PN) \cite{Blanchet:2013haa,Schafer:2018jfw}, Multipolar-Post-Minkowskian (MPM) \cite{Blanchet:1985sp,Blanchet:1989ki}, and gravitational self-force (GSF) \cite{Barack:2018yvs} calculations provide information about the evolution of the binary within different approximation schemes, such as the equations of motion, fluxes, orbital frequencies and radiative multipoles, that can be translated directly into a time-dependent waveform. Similarly, radiation-reaction effects, higher-order multipoles, nonlinear memory and other hereditary contributions have an intrinsically time-dependent structure \cite{Blanchet:1987wq,Blanchet:1992br,Blanchet:1997jj,Christodoulou:1991cr,Favata:2008yd,Talbot:2018sgr,Elhashash:2025hqi,Georgoudis:2025vkk,Bini:2026vaq}. In the time domain, these effects can be evolved together with the orbital dynamics without first requiring a transformation into a stationary frequency representation.
Furthermore, hereditary effects depend on the entire past history of the source and therefore contribute directly to the accumulated phase and amplitude of the time-domain signal.

Frequency-domain waveforms nevertheless play a central role in gravitational-wave data analysis, particularly for quasi-circular inspirals. Their principal advantage is computational: for slowly evolving systems, the Fourier transform can turn certain operations involving convolutions and differential equations into algebraic relations, and frequency-domain templates can be generated and evaluated very efficiently. This has made them extremely useful for matched-filter searches and parameter estimation \cite{Sathyaprakash:1991mt,LIGOScientific:2019hgc}. 
However, the computational efficiency of many analytic frequency-domain models relies on approximations such as the stationary-phase approximation \cite{Droz:1999qx,Yunes:2009yz}, whose underlying slowly evolving, locally monochromatic picture becomes increasingly inadequate as the binary approaches merger.

The time-domain representation is consequently particularly advantageous when the aim is to construct a waveform that remains faithful across the entire inspiral and through the transition toward merger. It naturally accommodates the breakdown of adiabaticity, rapid changes in orbital frequency, precession, eccentricity, higher harmonics and nonlinear radiation. It also avoids the need to reconstruct the temporal structure of the signal from a frequency-domain approximation whose validity may vary across the parameter space. For this reason, time-domain waveforms provide an especially natural interface between analytical calculations of the two-body dynamics and numerical-relativity simulations, and they are well suited to effective-one-body (EOB) \cite{Buonanno:1998gg,Damour:2000we} constructions in which the information from different approaches is combined with strong-field numerical data.

Although astrophysical gravitational-wave sources are usually described as bound systems, the scattering problem contains much of the same information about the underlying two-body dynamics. 
During a gravitational scattering event, the accelerated bodies emit gravitational radiation, producing a burst-like signal that is naturally described as gravitational bremsstrahlung \cite{Kocsis:2006hq,Berry:2012im,Mukherjee:2020hnm}. In contrast to a quasi-circular inspiral, where the signal is characterized by a slowly evolving carrier frequency, a scattering event is intrinsically non-adiabatic: the orbital frequency changes rapidly near closest approach and the radiation is localized around the encounter. The waveform therefore contains direct information about the time-dependent acceleration of the bodies and about nonlinear gravitational interactions.

In the case of a scattering configuration the post-Minkowkian (PM) approach is especially convenient because the bodies can be treated as asymptotically free in the remote past and future, while their interaction is expanded systematically in powers of Newton's constant (see, e.g., Refs. \cite{Damour:2016gwp,Damour:2017zjx,Damour:2019lcq} ad references therein). At leading order, the bodies follow approximately straight-line trajectories and exchange a single graviton; higher PM orders describe successive gravitational interactions and nonlinear corrections. The resulting scattering angle, impulse, radiated momentum and energy, and gravitational waveform provide gauge-invariant observables that characterize the relativistic two-body interaction.
They are particularly useful for matching calculations performed in different frameworks. 
New approaches have rapidly developed in recent years based on the connection between classical gravitational dynamics and quantum-field methods established by the effective field theory (EFT) description of compact binaries \cite{Goldberger:2004jt,Foffa:2013qca,Porto:2016pyg,Levi:2018nxp,Foffa:2016rgu,Foffa:2019hrb,Bjerrum-Bohr:2018xdl,Kosower:2018adc,Cristofoli:2021vyo,Cheung:2018wkq,Bjerrum-Bohr:2019kec,Bern:2019crd,Bern:2019nnu,Mogull:2020sak,Kalin:2020mvi,Kalin:2020fhe,Bern:2021dqo,Bern:2021yeh,Dlapa:2021npj,Dlapa:2021vgp,Bjerrum-Bohr:2021din,Saketh:2021sri,Kalin:2022hph,Khalil:2022ylj,Herrmann:2021lqe,Herrmann:2021tct,Mougiakakos:2021ckm,Jakobsen:2021smu,Riva:2021vnj,Manohar:2022dea,DiVecchia:2021bdo,DiVecchia:2023frv,Damgaard:2023ttc,Georgoudis:2023eke,Georgoudis:2024pdz,Heissenberg:2025ocy,Heissenberg:2025fcr,Dlapa:2022lmu,Driesse:2024xad,Driesse:2024feo,Porto:2024cwd,Dlapa:2024cje,Dlapa:2025biy,Dlapa:2026oyq,Brunello:2025gpf,Brunello:2026anu}. 
Modern amplitude techniques make it possible to extract classical gravitational observables from quantum-field scattering amplitudes by isolating the contributions that survive in the classical limit. The scattering observables can then be related to tree-level and loop amplitudes, providing computational tools that are often substantially different from direct perturbative solutions of Einstein's equations.

The scattering angle is particularly important because it is an asymptotic observable encoding the dynamics, and can therefore be calculated independently using different theoretical descriptions. 
Such an interplay becomes even more powerful when radiation is included. 
The radiated four-momentum, provides an additional observable that complements the conservative scattering angle and allows the dissipative sector of the two-body dynamics to be studied within the same framework.
The PM treatment of radiation reaction has reached 4PM order, including linear, nonlinear and hereditary radiation-reaction effects \cite{Bern:2021dqo,Bern:2021yeh,Dlapa:2022lmu}. Recent works have also initiated the study of the 5PM scattering dynamics in the first-order self-force (1SF) limit and beyond \cite{Driesse:2024xad,Driesse:2024feo,Dlapa:2025biy,Dlapa:2026oyq,Driesse:2026qiz}.

The gravitational bremsstrahlung waveform has been recently computed in the MPM framework at 4PM order and 2PN order in Ref. \cite{Bini:2024ijq}, providing an explicit frequency-domain waveform and its radiative multipole decomposition.
Amplitude-based calculations have now reached a level at which they can be compared directly with established PN and MPM results.
The current knowledge of the scattering waveform is at the one-loop level \cite{Brandhuber:2023hhy,Herderschee:2023fxh,Elkhidir:2023dco,Georgoudis:2023lgf}, and agrees with the MPM one up to the 2.5PN order, which is the level of accuracy reached in Ref. \cite{Bini:2023fiz} (see also Ref. \cite{Bini:2024rsy}).
Further agreement at the 3.5PN order has also been discussed in Refs. \cite{Bini:2026dvn,Bini:2026jwc} for the even-parity quadrupolar and octupolar parts of the waveform, respectively.
Notably, EFT-inspired methods have recently allowed for the computation of the leading-order (tree-level) time-domain waveform for spinless bodies \cite{Jakobsen:2021smu,Mougiakakos:2021ckm}, reproducing the seminal Kovacs and Thorne result \cite{Kovacs:1977uw,Kovacs:1978eu} obtained in the framework of classical perturbation theory, later extended to the spinning case in Refs. \cite{Jakobsen:2021lvp,DeAngelis:2023lvf}.

In previous works \cite{Geralico:2026kbm,Geralico:2026efi} I computed the frequency-domain bremsstrahlung waveform at the 8PM (six-loop) level with 6PN accuracy to first order in the binary's mass ratio, i.e., within the 1SF approximation.
According to the GSF theory, in the extreme-mass-ratio limit the smaller body follows a geodesic of the background spacetime generated by the larger object at zeroth order, while its own gravitational field perturbs the background field at successive orders in the mass ratio.
The underlying formalism is well established, and 1SF calculations are now capable of determining highly accurate conservative and dissipative corrections to both bound and unbound observables, such as redshift invariant, periastron advance, scattering angle, and radiative losses.
The significance of GSF theory actually extends well beyond the extreme-mass-ratio regime for which it was originally developed. 
In fact, self-force calculations provide gauge-invariant information about the two-body dynamics that can be translated into PN and EOB potentials, allowing for nontrivial cross-checks. In particular, 1SF results can constrain the terms linear in the symmetric mass ratio in EOB potentials, while self-force information at high PN order can be used to identify or validate coefficients in the PN expansions of PN-based models, e.g., the Tutti Frutti approach \cite{Bini:2019nra,Bini:2020hmy,Bini:2020nsb,Bini:2020wpo,Bini:2020rzn,Bini:2025zvk}. Conversely, high-order PN calculations provide independent checks of self-force calculations in their common domain of validity.

In this work I provide the 1SF time-domain scattering waveform at the 5PM (three-loop) level with 7PN accuracy, used in Ref. \cite{Geralico:2025rof} to compute the 5PM radiated energy and 4PM radiated angular momentum with the same PN accuracy.
The former was found to agree with the result of Ref. \cite{Driesse:2024feo}, whereas the latter has been recently confirmed by Ref. \cite{Heissenberg:2025fcr} by using amplitude-based methods, thereby providing an independent check.
In the extreme mass-ratio limit the waveform can be computed in the framework of first-order perturbation theory following the standard Teukolsky formalism.
The smaller mass $m_1$ is assumed to move along a hyperboliclike geodesic orbit on the equatorial plane of a Schwarzschild spacetime (with mass $m_2$).
All the information on the radiation emitted by the system is encoded in the Weyl scalar $\psi_4$, which is asymptotically related to the waveform of the emitted gravitational waves.

Units are chosen so that $G=1=c$, unless otherwise specified.
The standard PN small parameter $\eta=1/c$ is used to keep track of the order of the PN expansion of the various quantities.

\section{Reminder on hyperbolic-like geodesics in a Schwarzschild background}

Consider a Schwarzschild spacetime with line element $ds^2=\bar g_{\alpha\beta}dx^\alpha dx^\beta$ written in standard spherical-like coordinates $(t,r,\theta,\phi)$ as
\beq
ds^2 =-fdt^2 + \frac{1}{f}dr^2 + r^2(d\theta^2+\sin^2\theta d\phi^2)\,,
\eeq
where $f=1-\frac{2m_2}{r}$.
A hyperboliclike orbit on the equatorial plane ($\theta=\frac{\pi}{2}$) is the timelike geodesic with parametric equations $x^\mu =x_p^{\mu}(\tau)$ and 4-velocity $\bar u=\bar u^\alpha\partial_\alpha=\frac{dx_p^\alpha}{d\tau}\partial_\alpha$, such that
\bea
\label{geoeqns}
\frac{dt_p}{d\tau}&=&\frac{E}{f(r_p)}\,,\nonumber\\
\frac{dr_p}{d\tau}&=&\epsilon_r\left[E^2-f(r_p)\left(1+\frac{L^2}{r_p^2}\right)\right]^{1/2}\,,\nonumber\\
\frac{d\phi_p}{d\tau}&=&\frac{L}{r_p^2}\,,
\eea
where $\epsilon_r=\pm1$ is a sign indicator keeping track of increasing/decreasing radial coordinate, and $E=-\bar u_t$ and $L=\bar u_\phi$ denote the particle's energy and angular momentum per unit mass, respectively.
For hyperbolic-like orbits $E>1$ and $L>L_{\rm crit}(E)$, where $L_{\rm crit}(E)$ is the critical value of the angular momentum for fixed energy corresponding to capture by the hole (see, e.g., Ref. \cite{Barack:2022pde}).
In this case the radial equation above admits three real roots, $r_1<0$, $r_2$ and $r_{\rm min}$, such that $2m_2<r_2<r_{\rm min}$, 
the latter denoting the closest approach distance.
$E$ and $L$ are related to the initial velocity at infinity $v$ and the impact parameter $b$ by
\beq
E=\frac{1}{\sqrt{1-v^2}}\,, \qquad
L=b\sqrt{E^2-1}=bvE\,,
\eeq 
with $0<v<1$ and $b>b_{\rm crit}=L_{\rm crit}(E)/(vE)$.
Weak-field (i.e., PM) solutions to the geodesic equations correspond to large values of the impact parameter, i.e., $m_2/b\ll1$, at fixed values of the velocity $v$. Small values of the velocity then gives the PN expansion.

Using $u$ as a parameter the geodesic equations \eqref{geoeqns} read 
\bea
\label{geoeqnsu}
\frac{dt_p}{du}&=&-\epsilon_r\frac{m_2\sqrt{1+2\bar E}}{ju^2(1-2u)}\left(2u^3-u^2+\frac{2u}{j^2}+\frac{2\bar E}{j^2}\right)^{-1/2}
\,,\nonumber\\
\frac{d\phi_p}{du}&=&-\epsilon_r\left(2u^3-u^2+\frac{2u}{j^2}+\frac{2\bar E}{j^2}\right)^{-1/2}\,,
\eea
where we have introduced the dimensionless energy and angular momentum parameters $\bar E$ and $j$ defined by $E=\sqrt{1+2\bar E}$ and $L=m_2j$, respectively.
The above equations can be solved in closed analytical form in terms of Elliptic integrals.
However, for our purposes it is enough to solve them by using a PN expansion ($m_2\to m_2\eta^2$, $u\to u\eta^2$, $j\to j/\eta$, $\bar E\to \bar E\eta^2$). 
Unfortunately, the solution $t_p(u)$ cannot be easily inverted even at the Newtonian level, due to the presence of nontrivial logarithmic functions (see Eq. (2.19) of Ref. \cite{Bini:2024icd}). 

In order to get an explicit solution of the geodesic equations which can be easily expanded in both PM and PN approximations it is convenient to introduce the polar representation of the orbit
\beq
r_p=\frac{m_2p}{1+e\cos \chi}\,,
\eeq
where $p$ is the (dimensionless) semi-latus rectum and $e$ the eccentricity, which are related to $E$ and $L$ by
\bea
E&=&\sqrt{\frac{(p-2)^2-4e^2}{p(p-3-e^2)}}\,,\nonumber\\
L&=&\frac{m_2p}{\sqrt{p-3-e^2}}\,, 
\eea
with $e>1$ and $p>6+2e$.
The relativistic anomaly $\chi$ takes values in the range $(-\chi_\infty,\chi_\infty)$, with $\chi_\infty={\rm arccos}(-1/e)$, the value $\chi=0$ corresponding to the minimum approach distance $r_{\rm min}=m_2p/(1+e)$.

In terms of the relativistic anomaly the geodesic equations \eqref{geoeqns} write as
\bea
\label{geoeqnschi}
\frac{dt_p}{d\chi} &=& \frac{ m_2 p^2}{(p-2-2e\cos\chi)(1+e\cos\chi)^2} \sqrt{\frac{(p-2)^2-4e^2}{p-6-2e\cos\chi}}\,,\nonumber\\
\frac{d\phi_p}{d\chi} &=& \sqrt{\frac{p}{p-6-2e\cos\chi}}\,.
\eea
The PM expansion of the orbit can be obtained by expressing the orbital parameters $(p,e)$ in terms of $(b,v)$ in a weak-field limit (i.e., for large values of the impact parameter $b$) as follows (see Ref. \cite{Barack:2022pde} for details)
\bea
\label{epexp}
e&=&\frac{1}{\epsilon}\left[1-\frac{1}{2}(4v^2-1+8v^4)\epsilon^2
-\frac{1}{8}(64v^4-8v^2+256v^6+1)\epsilon^4+O(\epsilon^6)\right]
\,,\nonumber\\
\frac1p&=&v^2\epsilon^2\left[1+4v^2(1+v^2)\epsilon^2
+16v^4(2+4v^2+v^4)\epsilon^4+O(\epsilon^6)\right]
\,,
\eea
where the small PM parameter $\epsilon=\frac{m_2}{bv^2}\ll1$ has been introduced.
The solution for the relativistic anomaly in the large-$b$ expansion limit is given by
\bea
\chi(T)&=& {\rm arctan}(T)
+ \epsilon\left[\frac{T}{\sqrt{T^2+1}}+\frac{\left(1-3 v^2\right){\rm arcsinh}(T)}{\left(T^2+1\right)}\right]
+\epsilon^2\left[\frac{2Tv^2}{\left(T^2+1\right)}\right.\nonumber \\
&&\left.
-\frac{T \left(3 v^2-1\right)^2 {\rm arcsinh}^2(T)}{\left(T^2+1\right)^2}+\frac{\left(3 v^2-2\right) \left(3
   v^2-1\right){\rm arcsinh}(T)}{\left(T^2+1\right)^{3/2}}-\frac{15}{2}v^4\frac{{\rm arctan}(T)}{T^2+1}\right]
+O(\epsilon^3)\,,
\eea
where 
\beq
T=\frac{vt}{b} 
\eeq
is a rescaled (dimensionless) time.

The first few terms of the PM expansion of the orbit then read
\bea
\label{solgeoT}
u_p(T)&=&\epsilon\frac{v^2}{\sqrt{T^2+1}}
+\epsilon^2v^2\left[\frac{{\rm arcsinh}(T) \left(3 v^2-1\right)}{\left(T^2+1\right)^{3/2}}+\frac{1}{\left(T^2+1\right)}\right]
+\epsilon^3v^2\left[\frac{4 v^2+1}{2 \left(T^2+1\right)^{3/2}}\right.\nonumber \\
&&\left.
+\frac{\left(2 T^2-1\right) \left(3 v^2-1\right)^2 {\rm arcsinh}^2(T)}{2 \left(T^2+1\right)^{5/2}}-\frac{3 T \left(v^2-1\right) \left(3v^2-1\right) {\rm arcsinh}(T)}{\left(T^2+1\right)^2 v^4}+\frac{15}{2}v^4\frac{T{\rm arctan}(T)}{\left(T^2+1\right)^{3/2}}\right]
+O(\epsilon^4)
\,,\nonumber \\
\phi_p(T)&=& {\rm arctan}(T)
+\epsilon\left[\frac{T \left(v^2+1\right)}{\sqrt{T^2+1}}+\frac{\left(1-3 v^2\right) {\rm arcsinh}(T)}{\left(T^2+1\right)}\right]
+\epsilon^2\left[\frac{3 T v^2\left(v^2+4\right)}{4 \left(T^2+1\right)}\right.\nonumber \\
&&\left.
+\frac{3 v^2\left(\left(T^2-9\right) v^2+4(1+T^2)\right) {\rm arctan}(T)}{4 \left(T^2+1\right)}-\frac{T \left(1-3 v^2\right)^2 {\rm arcsinh}^2(T)}{\left(T^2+1\right)^2}+\frac{2 (1-v^2) \left(1-3 v^2\right) {\rm arcsinh}(T)}{\left(T^2+1\right)^{3/2}}\right]\nonumber \\
&&
+O(\epsilon^3)
\,.
\eea

Substituting the PM expansion \eqref{epexp} of the orbital parameters then gives the combined PM-PN expansion form of the solutions for the geodesic orbit used in the next sections, further taking the low-velocity limit by introducing the small velocity parameter $p_\infty$ according to $v=\eta p_\infty/\sqrt{1+\eta^2p_\infty^2}$.

It is useful to introduce the new (dimensionless) temporal variable
\beq
\label{ydef}
y=T+\sqrt{1+T^2}\,, \qquad T = \frac{y^2 - 1}{2y}\,,
\eeq
such that $y\in(0,\infty)$ for $T\in(-\infty,\infty)$, and
\bea
\label{solgeoy}
u_p(y)&=&\epsilon\frac{2yv^2}{1+y^2}
+\epsilon^2v^2\left[\frac{4y^2}{\left(y^2+1\right)^2}+\frac{4\left(3 v^2-1\right) \left(y^2-1\right) y^2 \ln(y)}{\left(y^2+1\right)^3}\right]
+O(\epsilon^3)
\,,\nonumber \\
\phi_p(y)&=& {\rm arctan}\left(\frac{y^2-1}{2y}\right)
+\epsilon\left[\frac{\left(v^2+1\right) (y^2-1)}{y^2+1}-\frac{4 \left(3 v^2-1\right) y^2 \ln(y)}{\left(y^2+1\right)^2}\right]
+O(\epsilon^2)
\,.
\eea

\section{Time-domain waveform modes}

In the framework of first-order perturbation theory the Weyl scalar $\psi_4$ directly relates to the gravitational waveform at a large distance from the source
\beq
\label{hdef}
\psi_4(r\to\infty)\sim-\frac12(\ddot h_+-i\ddot h_\times)\,,
\eeq
where $h_+$ and $h_\times$ are the two independent polarizations of the gravitational waves.
In the Teukolsky formalism $\psi_4$ is decomposed into frequency modes and angular modes using spin-weighted spherical harmonics of spin weight $s=-2$ as follows
\beq
\label{sep}
\psi_4= \frac1{r^4}\int\frac{d\omega}{2\pi}e^{-i\omega t}\sum_{lm}\,\,R_{lm\omega}(r)\,\, {}_{-2}Y_{lm}(\theta,\phi)\,,
\eeq
where the radial function $R_{lm\omega}(r)$ satisfies the inhomogeneous Teukolsky equation sourced by the particle's energy momentum tensor. 
The asymptotic solution representing purely outgoing waves is given by 
\beq
R_{lm\omega }(r\to\infty)\sim Z^\infty_{lm\omega} r^3e^{i\omega r_*}\,,
\eeq
where $r_*$ is the tortoise coordinate, and $Z^\infty_{lm\omega}$ is the amplitude
\beq
\label{Zinf}
Z^\infty_{lm\omega}=\frac{C^{\rm trans}_{lm\omega}}{W_{lm\omega}}\int_{2m_2}^\infty dr\frac{R^{\rm in}_{lm\omega}(r)T_{lm\omega}(r)}{\Delta^2}\,,
\eeq
with $\Delta=r(r-2m_2)$, $W_{lm\omega}$ the (constant) Wronskian, $C^{\rm trans}_{lm\omega}$ the (constant) transmission coefficient, and $T_{lm\omega}(r)$ the harmonic decomposition of the source term.
I refer to Ref. \cite{Sasaki:2003xr} for notation and conventions as well as for the definition of the various quantities. 

The asymptotic form of $\psi_4$ then turns out to be
\beq
\label{sep2}
\psi_4= \frac1{r}\sum_{lm}\int\frac{d\omega}{2\pi}Z^\infty_{lm\omega}e^{-i\omega t_r}\,{}_{-2}Y_{lm}(\theta,\phi)\,,
\eeq
where $t_r=t-r_*$ is the retarded time, and the final expression for the amplitude \eqref{Zinf} has the form
\beq
\label{Zinf2}
Z^\infty_{lm\omega}=\frac{C^{\rm trans}_{lm\omega}}{W_{lm\omega}}\int dt'e^{i(\omega t'-m\phi_p(t'))}F_{lm\omega}(r_p(t'))\,.
\eeq
The function $F_{lm\omega}(r_p(t))$ is the result of the action of a differential operator on the ingoing homogeneous solution $R^{\rm in}_{lm\omega}(r)$ upon evaluation at the particle position $r=r_p(t)$.
It can be formally written as follows
\beq
\label{Flmomega}
F_{lm\omega}(r_p(t))=\left[\alpha_{lm\omega}(r) R^{\rm in}_{lm\omega}(r) + \beta_{lm\omega}(r) \frac{dR^{\rm in}_{lm\omega}(r)}{dr}\right]_{r=r_p(t)}\,,
\eeq
with coefficients $\alpha_{lm\omega}$ and $\beta_{lm\omega}$ which also depend on the $\theta$-part of the spin-weighted spherical harmonic with given $lm$ and its derivative evaluated on the equatorial plane (the $\phi$-part having been factored out).

The time-domain waveform is defined as
\beq
h(t_r,\theta,\phi)=\lim_{r\to \infty}(r( h_+ -  i h_\times))\,,
\eeq
and is determined by the asymptotic form \eqref{sep2} of $\psi_4$, which yields 
\beq
\label{hdef}
h(t_r,\theta,\phi)=2G\sum_{lm}\int\frac{d\omega}{2\pi}\frac{Z^\infty_{lm\omega}}{\omega^2}e^{-i\omega t_r}\,{}_{-2}Y_{lm}(\theta,\phi)\,.
\eeq
It is useful to introduce a rescaled waveform 
\bea
\label{Wdef}
{\mathcal W}(t_r,\theta,\phi)&=&\frac{1}{4G}h(t_r,r,\theta,\phi)\nonumber\\
&=&\sum_{lm}{\mathcal W}_{lm}(t_r)\,{}_{-2}Y_{lm}(\theta,\phi)\,,
\eea
where ${\mathcal W}_{lm}(t_r)$ are the waveform modes in the time domain
\bea
{\mathcal W}_{lm}(t_r)&=&\int\frac{d\omega}{2\pi}e^{-i\omega t_r}\frac{C^{\rm trans}_{lm\omega}}{2\omega^2W_{lm\omega}}
\int dt'e^{i(\omega t'-m\phi_p(t'))}F_{lm\omega}(r_p(t'))
\nonumber\\
&\equiv&\int dt' K_{lm}(t',t_r)\,,
\eea
with
\bea
\label{Klm}
K_{lm}(t',t_r)&=&\int \frac{d\omega }{2\pi}e^{i\omega(t'-t_r)}\frac{C^{\rm trans}_{lm\omega}}{2\omega^2W_{lm\omega}} e^{-im\phi_p(t')} F_{lm\omega}(r_p(t'))
\nonumber\\
&\equiv&\int \frac{d\omega }{2\pi}e^{i\omega(t'-t_r)}K_{lm\omega}(t')\,.
\eea

The function $F_{lm\omega}(r_p(t'))$ given in Eq. \eqref{Flmomega} is computed by using the Mano, Suzuki and Takasugi (MST) \cite{Mano:1996mf,Mano:1996vt} homogeneous solutions to the radial Teukolsky equation satisfying the retarded boundary conditions of ingoing radiation at the horizon and upgoing at infinity.

The MST solutions are not simply polynomial functions of the frequency just as the PN homogeneous solutions, but also involve powers of logarithms and powers of $\omega$.
Substituting then in Eq. \eqref{Flmomega} implies that the function $K_{lm\omega}(t')$ in Eq. \eqref{Klm} has the general form
\beq
\label{Klmomega}
K_{lm\omega}(t')=\sum_{n,k\geq1} \left[c_{lmn}(t')+c^{\ln^k}_{lmn}(t')\ln^k\omega\right]\omega^n\,.
\eeq
Integration over frequencies is done straightforwardly in the case of polynomials $\omega^n$ in terms of the Dirac-delta function and its derivatives according to the rule $\omega^n\to i^{-n}\delta^{(n)}(t'-t)$.
Logarithmic terms $\omega^n\ln^k\omega$ instead lead to integrals of the type (see Appendix A)
\beq
I^{\ln}_k(t'-t_r)=\int \frac{d\omega }{2\pi}e^{i\omega(t'-t_r)}\ln^k\omega
=A_k(t_r-t')\frac{H(t_r-t')}{t'-t_r}+B_k\delta(t'-t_r)\,,
\eeq
where $H(x)$ denotes the Heaviside step function, and the coefficients $A_k(t_r-t')$ are polynomial functions of $\ln^{k-1}(t_r-t')$, whereas the $B_k$ are constants.
The function \eqref{Klm} thus reads
\beq
K_{lm}(t',t_r)=\sum_{n,k} \left[\left(c_{lmn}(t')+c^{\ln^k}_{lmn}(t')B_k\right)i^{-n}\delta^{(n)}(t'-t_r)
+(-i)^{-n}\frac{d^n}{dt^n}\left(c^{\ln^k}_{lmn}(t')A_k(t_r-t')\frac{H(t_r-t')}{t'-t_r}\right)\right]\,.
\eeq
Further integration over $t'$ then gives
\beq
{\mathcal W}_{lm}(t_r)=\sum_{n,k} (-i)^{-n}\frac{d^n}{dt_r^n}\left[\left(c_{lmn}(t')+c^{\ln^k}_{lmn}(t')B_k\right)\bigg\vert_{t'=t_r}
+\int_{-\infty}^{t_r}dt'\frac{c^{\ln^k}_{lmn}(t')A_k(t_r-t')}{t'-t_r}\right]\,.
\eeq
A more convenient way to compute the rescaled waveform modes consists in using the convolution theorem (see Appendix B), which yields 
\bea
\label{Wlmfin}
{\mathcal W}_{lm}(t_r)&=&\sum_{n,k} (-i)^{-n}\left[\frac{d^n}{dt'^n}\left(c_{lmn}(t')+c^{\ln^k}_{lmn}(t')B_k\right)\bigg\vert_{t'=t_r}
+\int_{-\infty}^{t_r}d\zeta\frac{A_k(t_r-\zeta)}{\zeta-t_r}\frac{d^nc^{\ln^k}_{lmn}(t')}{dt'^n}\bigg\vert_{t'=\zeta}\right]\nonumber\\
&\equiv&{\mathcal W}^{\rm inst}_{lm}(t_r)+{\mathcal W}^{\rm hered}_{lm}(t_r)\,,
\eea
where the time dependence of the coefficients $c_{lmn}(t')$ and $c^{\ln^k}_{lmn}(t')$ is only through $r_p(t')$ and $\phi_p(t')$, and I have distinguished an instantaneous part and a hereditary part.
The first term is straightforwardly computed by repeatedly using the geodesic equations \eqref{geoeqnsu} when taking the $n$-th derivative, i.e.,
\bea
\frac{du_p}{dt}&=&-\epsilon_r\frac{ju_p^2(1-2u_p)}{m_2\sqrt{1+2\bar E}}\left(2u_p^3-u_p^2+\frac{2u_p}{j^2}+\frac{2\bar E}{j^2}\right)^{1/2}
\,,\nonumber\\
\frac{d\phi_p}{dt}&=&\frac{ju_p^2(1-2u_p)}{m_2\sqrt{1+2\bar E}}\,,
\eea
to be PN-expanded through the replacements $m_2\to m_2\eta^2$, $u_p\to u_p\eta^2$, $j\to j/\eta$, $\bar E\to \bar E\eta^2$, with $2\bar E=p_\infty^2$ and $j=bp_\infty$.
One can then PM-expand the resulting expression by taking its large-$b$ limit.

To compute the integrals in the second term of Eq. \eqref{Wlmfin} instead one needs the explicit solution for the orbit as a function of time.
The integrals are of the type 
\beq
\label{nlint}
\int_{-\infty}^{t_r}dt'\,\frac{F(t')\ln^{k-1}(t_r-t')}{t'-t_r}\,,
\eeq
which diverge for $t'=t_r$, and are evaluated by taking their Hadamard finite part (see Appendix B).

\section{Results}

It is convenient to distinguish in both instantaneous and hereditary waveform modes an even-parity part and an odd-parity part 
\beq
{\mathcal W}_{lm}(t_r)={\mathcal W}^{\rm even}_{lm}(t_r)+{\mathcal W}^{\rm odd}_{lm}(t_r)\,,
\eeq
which are in direct correspondence with the mass-type and current-type radiative multipole moments, respectively.
This is achieved by expressing the spin-weighted spherical harmonics as linear combination of (the $\theta$-part of) ordinary spherical harmonics $Y_{lm}(\theta)$ (with the same $l$ and $m$) and their $\theta$-derivative $Y_{lm}'(\theta)$, so that the coefficients defining the function $F_{lm\omega}(r_p(t))$, Eq. \eqref{Flmomega}, writes as  
\bea
\alpha_{lm\omega}(r_p(t))&=&Y^*_{lm}\left(\frac{\pi}{2}\right) \alpha^{Y}_{lm\omega}(r_p(t))
+Y_{lm}'{}^*\left(\frac{\pi}{2}\right) \alpha^{Y'}_{lm\omega}(r_p(t))\nonumber\\
&\equiv&\alpha^{\rm even}_{lm\omega}(r_p(t))+\alpha^{\rm odd}_{lm\omega}(r_p(t))\,,
\eea
and analogously for $\beta_{lm\omega}$.

The explicit expressions of the rescaled waveform modes ${\mathcal W}_{lm}={\mathcal W}^{*}_{l-m}$ for $l=2,\ldots,7$ (and the relevant values of $m$) are given in an ancillary file up to $O(G^4)$, corresponding to the 5PM (three-loop) level for the time-domain waveform $h(t_r,\theta,\phi)$.
The combined PM-PN expansion is done as a power series expansion in both parameters $Gm_2/b$ (labeling the PM order) and $p_\infty$ (each power of which corresponding to a half-PN order).
The instantaneous part is accurate to the 7PN order, whereas the hereditary part is computed with less PN accuracy at $O(G^4)$ due to the presence of nonlocal integrals which cannot be evaluated in terms of elementary functions as well as ordinary polylogarithms only, but also involve generalized polylogarithms.

Consider for instance the 1SF even-parity quadrupolar modes ${\mathcal W}^{\rm even}_{2m}$.
At $O(G^0)$ the only contribution is instantaneous, and represents the linear-in-$G$ stationary part of the time-domain waveform.
For $m=2$ one has 
\beq
{\mathcal W}^{{\rm even},\,G^0}_{22}=-\sqrt{5\pi}\left[
\frac{1}{5}p_\infty^2
-\frac{1}{14}p_\infty^4
+\frac{1}{24}p_\infty^6
-\frac{5}{176}p_\infty^8
+\frac{35 }{1664}p_\infty^{10}
-\frac{21 }{1280}p_\infty^{12}
+\frac{231 }{17408}p_\infty^{14}
-\frac{429 }{38912}p_\infty^{16}
+O(p_\infty^{18})\right]
\,,
\eeq
which can be resummed as follows
\beq
{\mathcal W}^{{\rm even},\,G^0}_{22}=\frac{\sqrt{5\pi}}{8 p_\infty^2}\left[
\left(3-2 p_\infty^2\right) \sqrt{p_\infty^2+1}-\frac{3}{p_\infty} {\rm arcsinh}(p_\infty)
\right]\,,
\eeq
and similarly for other constant modes.
The beginning of the PM-PN expansions of the instantaneous and hereditary parts are given by
\bea
\label{W22inst}
{\mathcal W}^{\rm even,\,inst}_{22}&=&-\sqrt{\frac{\pi}{5}}\bigg\{
p_\infty^2\left[1+O(p_\infty^2)\right]
+\frac{Gm_2^2}{b}\left[-\frac{2 i (y+i) \left(-y^2+3 i y+1\right)}{(-y+i)^3}
+O(p_\infty^2)\right]\nonumber\\
&+&
\frac{Gm_2^2}{b}\left(\frac{Gm_2}{bp_\infty^2}\right)\left[\frac{2 \left(-y^4+2 i y^3-4 y^2-2 i y-1\right)}{(-y+i)^4}-\frac{4 y^2 \left(-y^2+8 i y+1\right) \log (y)}{(-y+i)^5 (y+i)}
+O(p_\infty^2)\right]\nonumber\\
&+&
\frac{Gm_2^2}{b}\left(\frac{Gm_2}{bp_\infty^2}\right)^2\left[
\frac{2 i \left(-y^6+2 i y^5-3 y^4-10 i y^3+3 y^2+2 i y+1\right)}{(-y+i)^5 (y+i)}\right.\nonumber\\
&-&\left.
\frac{8 i \left(-y^2+8 i y+1\right) \left(y^2+3 i y-1\right) y^2 \log (y)}{(-y+i)^6 (y+i)^2}+\frac{8 \left(-y^4+10 i y^3-12 y^2-10 i y-1\right) y^3 \log ^2(y)}{(-y+i)^7 (y+i)^3}
+O(p_\infty^2)\right]\nonumber\\
&+&
\frac{Gm_2^2}{b}\left(\frac{Gm_2}{bp_\infty^2}\right)^3\left[
-\frac{2 (y-1) (y+1) \left(-y^6+2 i y^5-6 y^4+4 iy^3+6 y^2+2 i y+1\right)}{(-y+i)^6 (y+i)^2}\right.\nonumber\\
&+&
\frac{8 \left(-y^2+8 i y+1\right) \left(y^2+3 i y-1\right)^2 y^2 \log (y)}{(-y+i)^7 (y+i)^3}\nonumber\\
&+&
\frac{8 i \left(-2 y^6+9 i y^5-122 y^4-162 i y^3+122 y^2+9 i y+2\right) y^3 \log^2(y)}{(-y+i)^8 (y+i)^4}\nonumber\\
&-&\left.
\frac{16 \left(-y^6+12 i y^5-31 y^4-56 i y^3+31 y^2+12 i y+1\right) y^4 \log^3(y)}{(-y+i)^9(y+i)^5}
+O(p_\infty^2)\right]
+O(G^5)\bigg\}\,,
\eea
and 
\bea
\label{W22hered}
{\mathcal W}^{\rm even,\,hered}_{22}&=&-\sqrt{\frac{\pi}{5}}p_\infty^3\bigg\{
\frac{Gm_2^2}{b}\left(\frac{Gm_2}{bp_\infty^2}\right)\left[
-\frac{8 i \left(-2 y^2+11 i y+1\right) y^2}{(-y+i)^5}\right.\nonumber\\
&-&\left.
\frac{8 \left(-y^2+8 i y+1\right) y^2}{(-y+i)^5 (y+i)} 
\left(\ln\left(\frac{(y^2 + 1)^2}{y}\right) - \ln\left(\frac{2p_\infty}{b}\right)\right)
+O(p_\infty^2)\right]\nonumber\\
&+&
\frac{Gm_2^2}{b}\left(\frac{Gm_2}{bp_\infty^2}\right)^2\left[
\frac{16 i \left(-y^2+10 i y+1\right)y^2}{(-y+i)^6}
+\frac{16 i\left(-y^2+8 i y+1\right) \left(y^2+3 i y-1\right) y^2}{(-y+i)^6 (y+i)^2}\ln\left(\frac{2p_\infty}{b}\right)\right.\nonumber\\
&+&
\frac{16 i y^2 \left(-y^6+2 i y^5-57 y^4-68 i y^3+57 y^2+2 i y+1\right) \log (y)}{(-y+i)^7 (y+i)^3}\nonumber\\
&+&
\frac{16 y^3 \left(-y^4+10 i y^3-12 y^2-10 i y-1\right)}{3 (-y+i)^7 (y+i)^3} 
\left(6\log (y)\left(\ln\left(\frac{(y^2 + 1)^2}{y}\right) - \ln\left(\frac{2p_\infty}{b}\right)\right)+6 {\rm Li}_2(-y^2)-\pi ^2\right)\nonumber\\
&+&\left.
O(p_\infty^2)\right]\nonumber\\
&+&
\frac{Gm_2^2}{b}\left(\frac{Gm_2}{bp_\infty^2}\right)^3\left[
\left(-\frac{96 y^4 \left(-y^6+12 i y^5-31y^4-56 i y^3+31 y^2+12 i y+1\right) \log ^2(y)}{(-y+i)^9 (y+i)^5}\right.\right.\nonumber\\
&+&\left.
\frac{16 y^2 \left(-y^2+8 i y+1\right) \left(y^2+3 i y-1\right)^2}{(-y+i)^7 (y+i)^3}\right)
\left(\ln\left(\frac{(y^2 + 1)^2}{y}\right) - \ln\left(\frac{2p_\infty}{b}\right)\right)\nonumber\\
&-&
\frac{16 i y^3 \left(-2 y^6+9 i y^5-122 y^4-162 i y^3+122 y^2+9 i y+2\right)}{3(-y+i)^8 (y+i)^4} 
\left(6 \ln\left(\frac{2p_\infty}{b}\right) \log (y)+6 {\rm Li}_2(-y^2)+\pi ^2\right)\nonumber\\
&+&
\frac{96 y^4 \left(-y^6+12 i y^5-31 y^4-56 i y^3+31 y^2+12 i y+1\right)}{(-y+i)^9 (y+i)^5}
\left(\left(\frac{\pi^2}{3}-2{\rm Li}_2(-y^2)\right) \log (y)+2{\rm Li}_3(-y^2)-2\zeta(3)\right)\nonumber\\
&+&
\frac{16 y^3 \left(-11 y^7+144 i y^6-473 y^5-1036 i y^4+779 y^3+392 i y^2-7 y+4 i\right) \log ^2(y)}{(-y+i)^9 (y+i)^5}\nonumber\\
&-&
\frac{32 i y^3 \left(-3 y^6+22 i y^5-155 y^4-220 i y^3+155 y^2+22 i y+3\right) \log (y)}{(-y+i)^8 (y+i)^4}\nonumber\\
&+&\left.
\frac{8 i y^2 \left(-2 y^7+7 i y^6-16 y^5-54 i y^4+54 y^3+61 i y^2-12 y+2 i\right)}{(-y+i)^7 (y+i)^3}
+O(p_\infty^2)\right]
+O(G^5)\bigg\}\,,
\eea
respectively.

\section{Conclusions}

I provide here the expression for the 5PM (three-loop) time-domain scattering waveform of two nonspinning bodies to first order in their mass ratio, recently used in Ref. \cite{Geralico:2025rof} to compute the 5PM radiated energy and 4PM radiated angular momentum with the 7PN accuracy.
This result is complementary to the frequency domain-waveform I computed in previous works through a high PM order.
In fact, although frequency-domain representations are extremely efficient for many data-analysis applications, the time-domain description retains a direct correspondence with both the instantaneous binary dynamics and hereditary effects induced by  radiation reaction as well as nonlinear interactions.
The PM level reached by the present calculation is far beyond the leading order waveform of Kovacs and Thorne, which is also the state-of-the-art of current amplitude-based time-domain computations.
Therefore, even if limited to extreme-mass-ratio configurations the present result will be useful for the cross-checking of ongoing calculations by different approaches.

\section*{Acknowledgments}

I would like to thank Thibault Damour and Donato Bini for useful discussions at the early stage of the present project. 
I acknowledge the Istituto per le Applicazioni del Calcolo ``M. Picone,'' CNR, for past support and hospitality.
I'm grateful to Davide Usseglio for drawing my attention to a missing term in Eq. \eqref{W22hered}, which was copied incorrectly from the associated code.

\appendix

\section{Inverse Fourier transform of $\ln^k(\omega)$ integrals}

Consider the following integrals
\beq
\label{Ilnkdef}
I^{\ln}_k(t)=\int \frac{d\omega}{2\pi} e^{-i\omega t}\ln^k(\omega)\,,
\eeq
which enter the hereditary part of the waveform modes.
For $k=1,\ldots,4$ (which are enough for the reached PM-PN accuracy) one has
\bea
\label{Ilnkfin}
I^{\ln}_1(t)&=& -\frac{1}{t}H(t)-\left(\gamma -i\frac{\pi}{2} \right)\delta(t)\,,\nonumber\\
I^{\ln}_2(t)&=& \frac{ 2\gamma  -i\pi + 2\ln t}{t} H(t)+\left(\gamma^2 -i\pi\gamma - \frac{5}{12}\pi^2  \right)\delta(t)\,,\nonumber\\
I^{\ln}_3(t)&=&\frac{-12\gamma^2 + 12i\gamma \pi+  5 \pi^2+ (-24\gamma + 12i\pi)\ln(t) - 12\ln^2(t)}{4 t} H(t)\nonumber\\
&+&\left(-\gamma^3 + \frac32 i \pi\gamma^2  + \frac54\pi^2\gamma-\frac{3i\pi^3}{8}  - 2 \zeta (3)\right)\delta(t)\,,\nonumber\\
I^{\ln}_4(t)&=&\frac{1}{2t}[8\gamma^3 - 12i\gamma^2\pi- 10\gamma \pi^2 + 3 i\pi^3 + 2 (12 \gamma^2 - 12i \gamma \pi - 5 \pi^2) \ln (t)\nonumber\\
&+&  12 (2 \gamma - i\pi )\ln^2 (t) + 8 \ln^3(t) + 16 \zeta(3)]H(t)\nonumber\\
&+& \left(\gamma^4 - 2i\pi\gamma^3  - \frac52\gamma^2 \pi^2 + 
 \frac32i\gamma\pi^3+\frac{79\pi^4}{240}+4(2\gamma-i\pi) \zeta(3) \right)\delta (t)\,,
\eea
as it is shown below.

Let us consider the principal branch of the complex logarithm $\log(\omega+i0)=\ln|\omega|+i\pi H(-\omega)$, where $H(x)$ denotes the Heaviside step function.
The integral $I^{\ln}_k(t)$ does not exist as an ordinary Lebesgue integral, but has to be interpreted as a tempered distribution.
Introduce then the complex parameter $\alpha$ and define
\beq
\label{Falpha}
F_\alpha(t)=\int \frac{d\omega}{2\pi} (\omega+i0)^{\alpha} e^{-i\omega t}\,,
\eeq
so that
\beq
\label{Ilnkdef2}
I^{\ln}_k(t)=\frac{\partial^k}{\partial\alpha^k}F_\alpha(t)\bigg\vert_{\alpha=0}\,.
\eeq
For $\omega>0$ $(\omega+i0)^{\alpha}=\omega^{\alpha}$, whereas for $\omega<0$ $(\omega+i0)^{\alpha}=|\omega|^{\alpha}e^{i\pi\alpha}$, so that 
\beq
F_\alpha(t)=\int_0^{\infty} \frac{d\omega}{2\pi} \omega^{\alpha} e^{-i\omega t}
+e^{i\pi\alpha}\int_0^{\infty} \frac{d\omega}{2\pi} \omega^{\alpha} e^{i\omega t}\,.
\eeq
For $t>0$ the standard analytically-continued Fourier-Mellin integral gives
\beq
\int_0^{\infty} d\omega\, \omega^{\alpha} e^{\pm i\omega t}=e^{\pm i\frac{\pi}{2}(\alpha+1)}\frac{\Gamma(\alpha+1)}{t^{\alpha+1}}\,,
\eeq
implying that
\beq
F_\alpha(t)=\frac{i}{\pi}e^{i\pi\alpha}\frac{\Gamma(\alpha+1)}{t^{\alpha+1}}\,, 
\qquad t>0\,.
\eeq
For $t<0$ the corresponding expression vanishes after analytical continuation.
Thus the distribution has support on the positive half-line.

It is convenient to write the result globally as
\beq
\label{Falphanew}
F_\alpha(t)=\frac{e^{i\frac{\pi}{2}\alpha}}{\Gamma(-\alpha)}\,t_+^{-\alpha-1}\,,
\eeq
where $t_+^\lambda=H(t)t^\lambda$ is understood as an analytically continued distribution.
For small values of $\alpha$ the series expansion of the prefactor is given by
\bea
\frac{e^{i\frac{\pi}{2}\alpha}}{\Gamma(-\alpha)}&=&-\alpha+\left(\gamma-i\frac{\pi}{2}\right)\alpha^2-\frac12\left(\gamma^2-\frac{5\pi^2}{12}-i\pi\gamma\right)\alpha^3+\left(\frac{\gamma ^3}{6}-\frac{1}{4} i \gamma ^2 \pi -\frac{5 \gamma  \pi ^2}{24}+\frac{i \pi ^3}{16}+\frac{\zeta (3)}{3}\right)\alpha^4\nonumber\\
&&
+\left(-\frac{\gamma  \zeta (3)}{3}+\frac{i \pi  \zeta (3)}{6}-\frac{\gamma ^4}{24}+\frac{1}{12} i \gamma ^3 \pi +\frac{5 \gamma ^2 \pi^2}{48}-\frac{1}{16} i \gamma  \pi ^3-\frac{79 \pi ^4}{5760}\right)\alpha^5
+O(\alpha^6)\,.
\eea
The distribution $t_+^{-\alpha-1}$ has a simple pole at $\alpha=0$. Its Laurent expansion is
\beq
t_+^{-\alpha-1}=-\frac1{\alpha}\delta(t)+\sum_{n=0}^{\infty}\frac{(-\alpha)^n}{n!}\frac{H(t)\ln^n t}{t}\,,
\eeq
so that 
\beq
t_+^{-\alpha-1}=-\frac1{\alpha}\delta(t)+\frac{H(t)}{t}-\alpha\frac{H(t)\ln t}{t}+\frac{\alpha^2}{2}\frac{H(t)\ln^2 t}{t}-\frac{\alpha^3}{6}\frac{H(t)\ln^3 t}{t}+\frac{\alpha^4}{24}\frac{H(t)\ln^4 t}{t}+O(\alpha^5)\,.
\eeq
Inserting the above expansions into Eq. \eqref{Falphanew} then gives
\bea
F_\alpha(t)&=&\delta(t)+\alpha\left[-\frac{H(t)}{t}-\left(\gamma-i\frac{\pi}{2}\right)\delta(t)\right]
+\alpha^2\left[\left(\gamma-i\frac{\pi}{2}+\ln t\right)\frac{H(t)}{t}+\frac12\left(\gamma^2-\frac{5\pi^2}{12}-i\pi\gamma\right)\delta(t)\right]\nonumber\\
&&
+\alpha^3\left[\left(-\frac{1}{2} \log ^2(t)+\left(-\gamma +\frac{i \pi }{2}\right) \log (t)+\frac{5 \pi ^2}{24}+\frac{i \gamma  \pi }{2}-\frac{\gamma ^2}{2}\right)\frac{H(t)}{t}\right.\nonumber\\
&&\left.
+\left(-\frac{\zeta (3)}{3}-\frac{\gamma ^3}{6}+\frac{1}{4} i \gamma ^2 \pi +\frac{5 \gamma  \pi ^2}{24}-\frac{i \pi ^3}{16}\right)\delta(t)\right]\nonumber\\
\nonumber\\
&&
+\alpha^4\left[\left(\frac{\log ^3(t)}{6}+\left(\frac{\gamma }{2}-\frac{i \pi }{4}\right) \log ^2(t)+\left(\frac{\gamma ^2}{2}-\frac{i \gamma  \pi }{2}-\frac{5 \pi^2}{24}\right) \log (t)+\frac{\zeta (3)}{3}+\frac{i \pi ^3}{16}-\frac{5 \gamma  \pi ^2}{24}-\frac{1}{4} i \gamma ^2 \pi +\frac{\gamma ^3}{6}\right)\frac{H(t)}{t}\right.\nonumber\\
&&\left.
+\left(\frac{\gamma  \zeta (3)}{3}-\frac{i \pi  \zeta (3)}{6}+\frac{\gamma ^4}{24}-\frac{1}{12} i \gamma ^3 \pi -\frac{5 \gamma ^2 \pi^2}{48}+\frac{1}{16} i \gamma  \pi ^3+\frac{79 \pi ^4}{5760}\right)\delta(t)\right]
+O(\alpha^5)\,.
\eea
The results \eqref{Ilnkfin} then follow straightforwardly from Eq. \eqref{Ilnkdef2}.

\section{Computation of divergent integrals}

Consider the function $K_{lm\omega}(t')$, Eq. \eqref{Klmomega}, namely
\beq
K_{lm\omega}(t')=\sum_{n,k\geq1} \left[c_{lmn}(t')+c^{\ln^k}_{lmn}(t')\ln^k\omega\right]\omega^n\,,
\eeq
to be further integrated over $t'$.
Integration over frequencies of the first term is done straightforwardly in terms of the Dirac-delta function and its derivatives by using
\beq
\int \frac{d\omega }{2\pi}e^{i\omega(t'-t_r)}\omega^n=i^{-n}\frac{d^n}{dt'^n}\int \frac{d\omega }{2\pi}e^{i\omega(t'-t_r)}
=i^{-n}\delta^{(n)}(t'-t_r)\,.
\eeq

The second term can be computed by using the convolution theorem, i.e., 
\beq
\int \frac{d\omega }{2\pi}e^{i\omega \xi}\hat f(\omega)\hat g(\omega)=\int d\tau g(\tau)f(\xi-\tau)
=\int d\tau f(\tau)g(\xi-\tau)\,,
\eeq
with $\hat f(\omega)=\omega^n$ and $\hat g(\omega)=\ln^k\omega$.
Consider, for instance, the case $k=1$.
We have
\bea
f(\xi)=i^{-n}\delta^{(n)}(\xi)\,, \qquad
g(\xi)=\frac{1}{\xi}H(-\xi)-\left(\gamma-\frac{i\pi}{2}\right)\delta(\xi)\,,
\eea
so that
\beq
c^{\ln}_{lmn}(t')\int \frac{d\omega }{2\pi}e^{i\omega(t'-t_r)}\omega^n\ln\omega=
c^{\ln}_{lmn}(t')\int d\tau\, i^{-n}\delta^{(n)}(t'-t_r-\tau)\left[\frac{1}{\tau}H(-\tau)-\left(\gamma-\frac{i\pi}{2}\right)\delta(\tau)\right]\,.
\eeq
Integrating over $t'$ then gives
\bea
\int dt'\,c^{\ln}_{lmn}(t')\int \frac{d\omega }{2\pi}e^{i\omega(t'-t_r)}\omega^n\ln\omega&=&
i^{-n}\int d\tau\,\left[\frac{1}{\tau}H(-\tau)-\left(\gamma-\frac{i\pi}{2}\right)\delta(\tau)\right]\int dt'\, c^{\ln}_{lmn}(t')\delta^{(n)}(t'-t_r-\tau)\nonumber\\
&=&i^{-n}\int d\tau\,\left[\frac{1}{\tau}H(-\tau)-\left(\gamma-\frac{i\pi}{2}\right)\delta(\tau)\right](-1)^n\frac{d^nc^{\ln}_{lmn}(t')}{dt'^n}\bigg\vert_{t'=t_r+\tau}\nonumber\\
&=&(-i)^{-n}\int d\tau\,\frac{1}{\tau}H(-\tau)\frac{d^nc^{\ln}_{lmn}(t')}{dt'^n}\bigg\vert_{t'=t_r+\tau}\nonumber\\
&&
-(-i)^{-n}\left(\gamma-\frac{i\pi}{2}\right)\frac{d^nc^{\ln}_{lmn}(t')}{dt'^n}\bigg\vert_{t'=t_r}
\,.
\eea
The first term can also be written as
\beq
(-i)^{-n}\int_{-\infty}^0 d\tau\,\frac{1}{\tau}\frac{d^nc^{\ln}_{lmn}(t')}{dt'^n}\bigg\vert_{t'=t_r+\tau}=
(-i)^{-n}\int_{-\infty}^{t_r} d\zeta\,\frac{1}{\zeta-t_r}\frac{d^nc^{\ln}_{lmn}(t')}{dt'^n}\bigg\vert_{t'=\zeta}\,,
\eeq
which diverges for $\tau=0$, and can be evaluated by taking its Hadamard's finite part.

The above integral is of the general form
\beq
\label{divint}
\int_{-\infty}^tdt'\frac{F(t')}{t'-t}\,,
\eeq
which is singular at $t'=t$.
One can regularize it by taking its Hadamard finite part 
\bea
{\rm Pf}_{\epsilon}\int_{-\infty}^tdt'\frac{F(t')}{t'-t}&=&\int_{-\infty}^tdt'\frac{F(t')-F(t)}{t'-t}+F(t)\int_{-\frac1{\epsilon}}^{t-\epsilon}\frac{dt'}{t'-t}\nonumber\\
&=&\int_{-\infty}^tdt'\frac{F(t')-F(t)}{t'-t}+2F(t)\ln\epsilon\,,
\eea
where the first term is now well behaved, and the parameter $\epsilon$ in the second term is formally considered infinitesimal, eventually replaced by a finite value playing the role of an arbitrary time scale.

In order to use the $y$-parametrization of the orbit one has to change variables as $t'\to y'=y(t')$ and $t\to y=y(t)$, so that 
\beq
\frac{dt'}{t'-t}=\frac{dy'}{t(y')-t(y)}\left(\frac{dt'}{dy'}\right)
=h(y',y)\frac{dy'}{y'-y}\,,
\eeq
with $h(y',y)=1+O(y'-y)$.
The integral \eqref{divint} thus becomes
\beq
\int_{-\infty}^tdt'\frac{F(t')}{t'-t}=\int_{0}^ydy'\frac{G(y',y)}{y'-y}\,,
\eeq
with $G(y',y)=h(y',y)F(y')$, so that 
\bea
{\rm Pf}_{\sigma}\int_{0}^ydy'\frac{G(y',y)}{y'-y}&=&\int_{0}^ydy'\frac{G(y',y)-G(y,y)}{y'-y}+F(y)\int_{0}^{y-\sigma(y)}\frac{dy'}{y'-y}\nonumber\\
&=&\int_{0}^ydy'\frac{G(y',y)-F(y)}{y'-y}+F(y)\ln\frac{\sigma(y)}{y}\,,
\eea
where $\sigma(y)=\epsilon f(y)$, $f(y)=\left(\frac{dt}{dy}\right)^{-1}$, is the new scale.
Recalling the definition of the time variable $y$, Eq. \eqref{ydef}, one has
\beq
t'-t = \frac{b}{v}\left(\frac{y'^2 - 1}{2y'}-\frac{y^2 - 1}{2y}\right)\,,
\eeq
and
\beq
h(y',y)= \frac{y(y'^2 + 1)}{y'(yy' + 1)}\,,
\eeq
with
\beq
\ln\sigma(y)=\ln(2) + \ln\epsilon+\ln(v)-\ln(b) + 2\ln(y) - \ln(y^2 + 1)\,.
\eeq

\end{document}